\documentclass[prl,twocolumn,floatfix]{revtex4}

\usepackage{graphicx}
\usepackage{bm}
\usepackage{amsmath}
\usepackage{color}
\definecolor{darkgreen}{rgb}{0.2,0.4,0.0}
\definecolor{darkblue}{rgb}{0.0,0.3,0.6}

\newcommand{\be}{\begin{equation}}
\newcommand{\ee}{\end{equation}}
\newcommand{\ba}{\begin{eqnarray}}
\newcommand{\ea}{\end{eqnarray}}
\newcommand{\nn}{\nonumber \\}

\begin{document}

      \title{Elimination of Perturbative Crossings in Adiabatic Quantum Optimization}
      \author{Neil G.~Dickson}
      \affiliation{D-Wave Systems Inc., 100-4401 Still Creek Drive,
      Burnaby, B.C., V5C 6G9, Canada}

\date{\today}

\begin{abstract}
It was recently shown that, for solving NP-complete problems, adiabatic paths always exist without finite-order perturbative crossings between local and global minima, which could lead to anticrossings with exponentially small energy gaps if present.  However, it was not shown whether such a path could be found easily.  Here, we give a simple construction that deterministically eliminates all such anticrossings in polynomial time, space, and energy, for any Ising models with polynomial final gap.  Thus, in order for adiabatic quantum optimization to require exponential time to solve any NP-complete problem, some quality other than this type of anticrossing must be unavoidable and necessitate exponentially long runtimes.
\end{abstract}
\maketitle

\section{Introduction}
The usefulness of adiabatic quantum optimization (AQO) for solving NP-complete problems, as originally proposed in \cite{Farhi01}, has been the subject of much debate in recent years. In AQO, the Hamiltonian of the quantum system begins as $H_B$, whose ground state is easy to prepare as the initial state of the system, and is evolved to a final Hamiltonian $H_P$, whose ground state is the optimal solution to an optimization problem. The running time, $t_f$, required to obtain a large amplitude of the ground state at the end of the evolution is proportional to $g^{-2}_{\rm min}$, where $g_{\rm min}$ is the minimum energy gap between the ground and first excited states, excluding excited states simply heading toward final degeneracy with the ground state, during the evolution. The complexity of AQO is therefore determined by the scaling of $g_{\rm min}$ with the problem size.

An example of an NP-hard \footnote{A problem is NP-hard if solving it in polynomial time allows all NP-complete problems to be solved in polynomial time.} optimization problem well suited to being represented by $H_P$ is that of finding the ground state of an Ising model with arbitrary couplings, $J_{ij}$, and arbitrary local fields, $h_i$:
\ba
H_P = \sum_i h_i \sigma_z^{(i)} + \sum_{i<j} J_{ij} \sigma_z^{(i)} \sigma_z^{(j)}.
\ea
It is easily seen that this $H_P$ is identical to the classical Ising model Hamiltonian, so its ground state is naturally identical to the ground state being sought (referred to here as the ``global minimum" to avoid confusion where applicable).  $H_B$ is often chosen to be of the form
\ba
H_B = -\sum_i \Delta_i \sigma_x^{(i)} \qquad \Delta_i>0,
\ea
whose ground state is the in-phase, uniform superposition of all Z-basis states.  If representations of $H_P$ and $H_B$ can be found for a problem such that $g_{\rm min}$ decreases only polynomially with the number of qubits, $n$, that problem can be solved in polynomial time using AQO.  This, of course, assumes that the quantum system used to implement AQO is sufficiently isolated from a thermal environment, but effects of such an environment are not examined here.

\begin{figure}[ht]
\includegraphics[trim=3.3cm 21.5cm 9.7cm 2.5cm,clip,width=8.5cm]{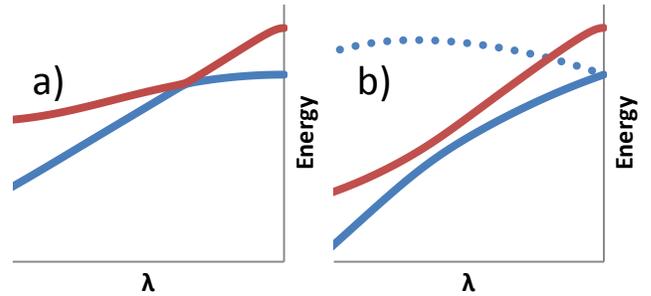}
\caption{a) The two lowest energy eigenstates in an AQO spectrum, which to low-order perturbation appear to cross, but actually anticross, causing a small $g_{\text{min}}$, and b) three lowest eigenstates after changing the final ground state to be twofold degenerate.  If the final ground state is degenerate, the corresponding eigenstates (solid and dotted blue) will repel each other away from the end.  If this degeneracy can be introduced without significantly affecting the excited state, the ground state energy (solid blue) will consequently be pushed away from that of the excited state (solid red).}
\label{Figure1}
\end{figure}

The complexity of AQO for solving NP-complete problems is a contentious issue, and has recently sparked intense debate.  It was originally conjectured that the gap size may scale polynomially with problem size \cite{Farhi01}, but counterexamples were found \cite{vanDam01,Znidaric06}, which were subsequently defeated by modifying the Hamiltonian \cite{Farhi02,Farhi05}.  Using perturbation expansion, Amin and Choi \cite{Amin09} showed that an eigenstate corresponding with a local minimum of $H_P$ (anti)crossing with that of the global minimum near the end of the evolution, sometimes called a first order quantum phase transition, can result in an exponentially small $g_{\rm min}$.  In particular, when a local minimum has more low-energy, possibly degenerate, states in its vicinity than the global minimum, the eigenstate corresponding to the local minimum may cross that of the global minimum.  Using the same perturbation argument, Altshuler {\em et al.}~\cite{Altshuler} showed that for one representation of random exact cover instances, the probability of having problematic crossings increases with the system size, and claimed that these crossings are unavoidable.  Others made similar observations using different techniques \cite{Young10,Jorg10}.

Others argued that the analysis in \cite{Altshuler} had not rigorously proven that the crossings are unavoidable, since problem structure that can impact the presence of such crossings was neglected \cite{Knysh10,Farhi09,Choi10,Dickson11}. Dickson and Amin~\cite{Dickson11} showed that there is always {\em some} selection of $H_B$ and $H_P$ that guarantees no crossings of local and global minima, but it was left open whether or not an {\em efficient} method of selecting the Hamiltonian exists.

Here, we present the first general method for eliminating perturbative crossings, based on the effects of degeneracy of the eigenstates of $H_P$, illustrated in Fig.~\ref{Figure1}.

\section{The Concept}
\begin{figure*}[ht!]
\includegraphics[trim=2.5cm 11.0cm 2.5cm 2.5cm,clip,width=17cm]{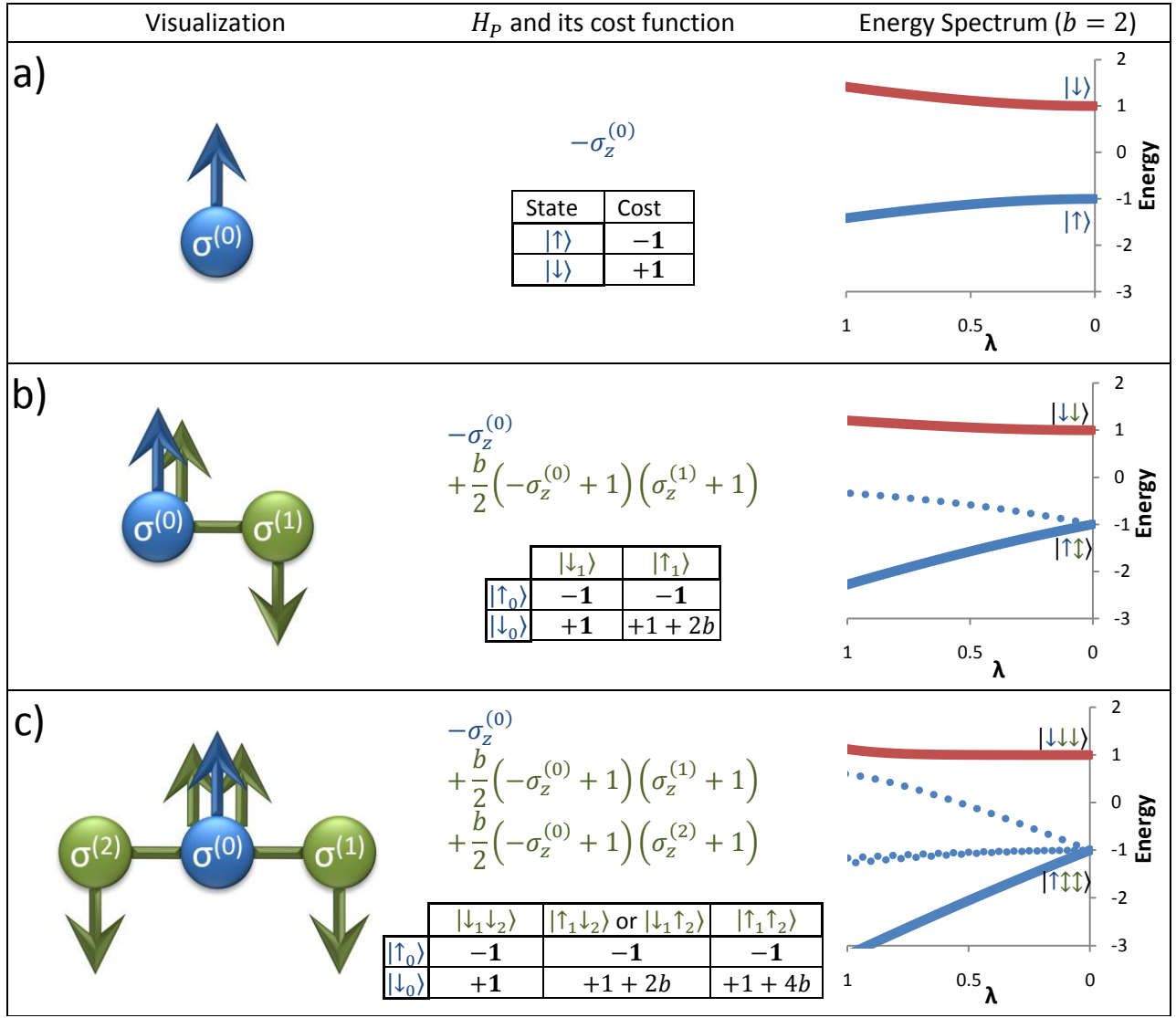}
\caption{Adding extra qubits in such a way that the ground state becomes degenerate, but the excited state remains nondegenerate, causes the final ground states to repel each other.  The ground state energy (solid blue) then diverges from the excited state energy (solid red).  Here, this is depicted for one original qubit (blue), and a) zero, b) one, and c) two extra qubits (green).  $\lambda$ is the perturbation parameter used in the analysis below.  For brevity, $\left|\textcolor{darkblue}{\uparrow}\right\rangle \otimes (\left|\textcolor{darkgreen}{\uparrow}\right\rangle + \left|\textcolor{darkgreen}{\downarrow}\right\rangle)/\sqrt{2}$ is written as $\left|\textcolor{darkblue}{\uparrow}\textcolor{darkgreen}{\updownarrow}\right\rangle$, and similar for $\left|\textcolor{darkblue}{\uparrow}\textcolor{darkgreen}{\updownarrow\updownarrow}\right\rangle$.  Note that the same method can be applied to a term $\textcolor{darkblue}{+\sigma_z^{(0)}}$ instead of $\textcolor{darkblue}{-\sigma_z^{(0)}}$ simply by substituting one for the other in $H_P$.  Likewise, the same method can be applied to a pair of original qubits coupled with a $J$ term $\textcolor{darkblue}{-\sigma_z^{(0)}\sigma_z^{(1)}}$ or $\textcolor{darkblue}{+\sigma_z^{(0)}\sigma_z^{(1)}}$, again by substituting over $\textcolor{darkblue}{-\sigma_z^{(0)}}$.}
\label{hTransform}
\end{figure*}

The main concept of the construction we present here is to force the ground state to diverge from all other states away from the end of the evolution by strategically introducing degeneracy, as depicted in Fig.~\ref{Figure1} and Fig.~\ref{hTransform}, eliminating any crossings that may have existed before applying the construction.  This can be done by adding extra qubits in such a way that the ground state becomes the most degenerate, the first excited state less degenerate, continuing this up to the highest excited state, which will be the least degenerate (possibly nondegenerate).  This degeneracy makes the eigenstates corresponding with the global minimum repel each other from the end more than all other eigenstates, so the ground state diverges from all other eigenstates as desired.  This must be done without any knowledge of the global minimum, apart from that it is the lowest energy state of $H_P$.

For simplicity, we will begin with the case of minimizing arbitrarily-connected Ising models where all $h,J \in \{-1,0,+1\}$, which is still NP-hard.  This can be easily generalized to arbitrary $h$ or $J$ values (of polynomial precision) without resorting to expensive reduction mappings onto $\{-1,0,+1\}$, as described later.  Even though the only structure in a general, dense Ising model is that at most 2 spins participate in each term of its Hamiltonian, this small amount of structure is enough to perform the required transformation.  This does not apply to random energy models, studied in \cite{Jorg10b,Farhi10}, since they have no applicable structure.

The construction introduces extra qubits for each nonzero $h$ or $J$ term of $H_P$ such that when the $h$ or $J$ term is satisfied (i.e. at its negative value), the extra qubits corresponding to that term are at degeneracy (i.e. flipping them does not change the energy of the system), and when the term is unsatisfied, the extra qubits are \emph{not} at degeneracy.  This is shown for an $h$ term in Fig.~\ref{hTransform}, and can be applied equivalently to a $J$ term.  Naturally, states with more terms satisfied will then be more degenerate than states with fewer terms satisfied, since states with more satisfied terms have more qubits at degeneracy.  In the case where $h,J \in \{-1,0,+1\}$, the global minimum has the most terms satisfied, and will thus have the most qubits at degeneracy.  As we will see, the slope ($1^{\text{st}}$-order perturbation energy correction) of a state will be proportional to the number of qubits at degeneracy in that state, and so the ground state will have the largest negative slope, diverging from all other states.

\section{The Construction}
As mentioned above, we will begin by examining the case of Ising models with $h,J \in \{-1,0,+1\}$.  The Hamiltonian of the original Ising model, $\textcolor{darkblue}{M}$, is given as

\ba
H_P = \sum_{i\in \textcolor{darkblue}{M}} \textcolor{darkblue}{h_i \sigma_z^{(i)}} + \sum_{\{i,j\}\in \textcolor{darkblue}{M}} \textcolor{darkblue}{J_{ij}\sigma_z^{(i)}\sigma_z^{(j)}}.
\ea

To introduce and amplify degeneracy of the ground state as desired, for each of the $m$ nonzero terms in $\textcolor{darkblue}{M}$, we need to be able to add an extra qubit such that when the term is satisfied, the energy is unchanged regardless of whether the extra qubit is $-1$ or $+1$.  However, when the term is unsatisfied, the energy must be unchanged when the extra qubit is $-1$, but higher by energy $2b>0$ when the extra qubit is $+1$.  This can be summarized as

\begin{center}
\begin{tabular}{| c || c | c |}
\hline
 & \multicolumn{2}{|c|}{{\bf Added energy cost}} \\
\hline
{\bf Term} & {\bf extra=$-1$} & {\bf extra=$+1$} \\
\hline
{\bf satisfied} & 0 & 0 \\
\hline
{\bf unsatisfied} & 0 & $2b$ \\
\hline
\end{tabular}
\end{center}

Note that if the energy didn't remain unchanged for some value of the extra qubit for both the satisfied and unsatisfied case, we would effectively be changing the energy of the original states, which is not what we intend, so all three zeros are important.  The added cost $2b$ serves only to break the degeneracy when the corresponding term is unsatisfied.

To construct the new terms to appear in $H_P$, one may first observe that the above table is simply a multiple of the truth table for a boolean AND operator, which is equivalent to binary variable multiplication, so we need only to convert our two conditions (``unsatisfied" and ``extra=+1") to binary variables and multiply by $2b$.  Since nonzero $h$ and $J$ are limited to $\pm 1$, this gives us $(\text{\bf term}+1)/2$ and $(\text{\bf extra}+1)/2$, so the term added to the Hamiltonian for the extra qubit should be
\ba
\textcolor{darkgreen}{b(\text{\bf term}+1)(\text{\bf extra}+1)/2}
\label{extraTerm}
\ea

This is illustrated in Fig.~\ref{hTransform}.  Now that we have a mechanism for adding these extra qubits, for each $h_i\ne 0$, add $a\ge 1$ extra qubits $\{i_1,...,i_a\}\in \textcolor{darkgreen}{M_h}$ with the following terms to the Hamiltonian. These are in exactly the same form as (\ref{extraTerm}).

\ba
&&\sum_{k=1}^{a} \textcolor{darkgreen}{b (h_i\sigma_z^{(i)}+1)(\sigma_z^{(i_k)}+1)/2}\nn
\ea

Similarly, for each $J_{ij}\ne 0$, add $a$ extra qubits $\{ij_1,...,ij_{a}\}\in \textcolor{darkgreen}{M_J}$ with the following terms to the Hamiltonian.

\ba
&&\sum_{k=1}^{a} \textcolor{darkgreen}{b(J_{ij}\sigma_z^{(i)}\sigma_z^{(j)}{+}1)(\sigma_z^{(ij_k)}{+}1)/2}
\ea

It is important to note that while this introduces 3-local (ZZZ) terms, equivalent, albeit much less intuitive, 2-local terms with the same property can be introduced by adding another extra qubit for each term.  This is described in the appendix below.

Since there are $m$ nonzero terms in the original Ising model, $\textcolor{darkblue}{M}$, we have added a total of $am$ extra qubits, each with an associated term coupling it to $\textcolor{darkblue}{M}$.

Special consideration must be made when $\textcolor{darkblue}{M}$ may already have single bit flip degeneracies.  The above construction still works for sufficiently (polynomially) large $a$ and $b$, but the analysis is made more complicated, so for simplicity of analysis in these cases, all qubits that may have a 1 bit flip degeneracy can be replaced by a strongly ferromagnetically coupled pair of qubits.  Since the pair is stongly coupled to align with each other, they will act logically as the original single qubit, any 1 bit flip degeneracy with a 2 bit flip degeneracy (i.e. both qubits in the pair must flip to remain at the same energy).  We are currently limiting the focus to $J$ values of $\pm 1$, but this strong ferromagnetism can be expressed with multiple (identical) $J=-1$ terms between the two qubits in the pair.  Thus, the analysis below safely assumes that there are no 1 bit flip degeneracies in the original Ising model.

\begin{figure*}[ht!]
\includegraphics[trim=2.5cm 10.6cm 2.5cm 2.5cm,clip,width=17cm]{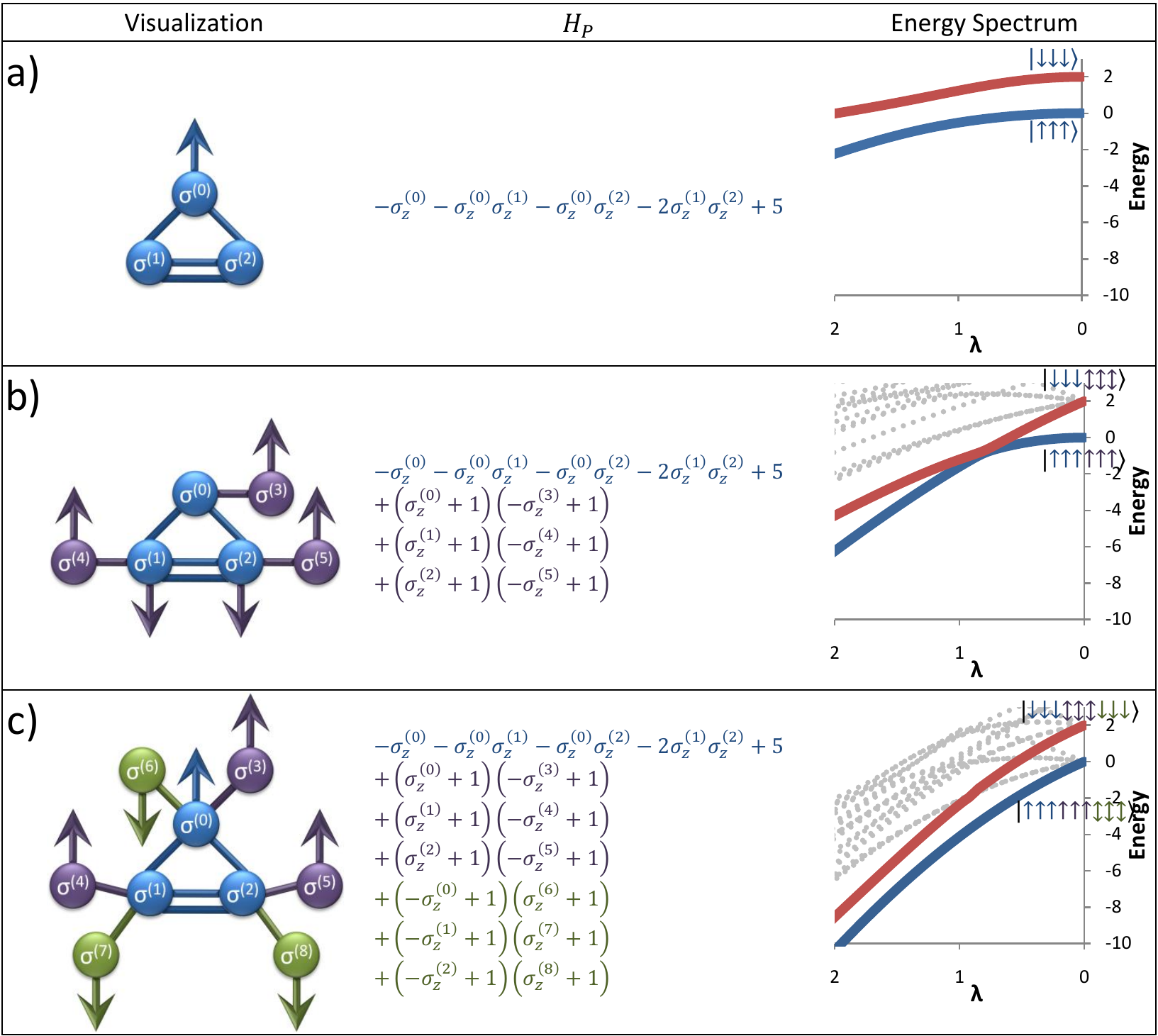}
\caption{a) Starting from a simple 3-qubit Ising model without a perturbative (anti)crossing, b) the known local minimum, $\left|\textcolor{darkblue}{\downarrow\downarrow\downarrow}\right\rangle$, can be made more degenerate (8-fold in this case) until a harsh crossing is created between the eigenstates corresponding with it and the global minimum, $\left|\textcolor{darkblue}{\uparrow\uparrow\uparrow}\right\rangle$.  c) Now that there is a crossing, it can be eliminated again by increasing the degeneracy of the known global minimum to match or preferably exceed that of the local minimum.  Although this does not apply the full construction (exact diagonalization is infeasible for the 42 qubits needed to eliminate the crossing), which does not require knowledge of the local or global minima, this example demonstrates the principle upon which the construction is based. (Note: For easier viewing, biases that exactly cancel are not depicted in the visualization.)}
\label{CreateEliminate}
\end{figure*}

Unfortunately, performing a full numerical simulation of this construction for relevant cases is infeasible, primarily due to the large number of added qubits necessary to apply it to Ising models sufficiently complex to have perturbative crossings.  Thus, it is examined only analytically below.  However, it is numerically feasible to demonstrate that changing the degeneracy, by adding extra qubits in a \emph{similar} manner, can be used to create and then eliminate these crossings.  An example is illustrated in Fig.~\ref{CreateEliminate}.

\section{Analysis}
We will be performing a perturbation analysis on the above construction by expressing the time-dependent Hamiltonian, $H$, as
\ba
\label{Hamiltonian}
H &=& H_P + \lambda H_B \nn
H_B &=& -\sum_i \Delta_i \sigma_x^{(i)},
\ea
where $\lambda$ is initially $\infty$, (though proportional to $abm$ suffices), and is monotonically decreased toward zero over time.  For now, we select all $\Delta_i=1$.

As mentioned above, the energies of the original states of $\textcolor{darkblue}{M}$ (with the additional qubits in their $-1$ states) are unchanged by the addition of the new qubits, and all added states with higher energy are 1 bit flip away from states with lower energy, and thus are not local minima.

Supposing first that $\textcolor{darkblue}{M}$ has a nondegenerate state $\alpha$, after applying the construction, $\alpha$ is now degenerate (unless of course all terms were unsatisfied, which means that $\alpha$ is the highest energy state).  In particular, upon introducing small $\lambda>0$, the lowest eigenstate of the degeneracy, $\left|\alpha\right>$, is always a uniform superposition of $\alpha$, since this uniquely minimizes $\left<\alpha\right|H_B\left|\alpha\right>$, the coefficient of the $1^{\text{st}}$-order term below.  We wish to examine the perturbation expansion of this eigenstate $\left|\alpha\right>$,
\ba
E_{\left|\alpha\right>}(\lambda) = E^{(0)}_{\left|\alpha\right>} + E^{(1)}_{\left|\alpha\right>} \lambda + E^{(2)}_{\left|\alpha\right>} \lambda^2 + ...
\ea

Because nonzero $h,J$ are limited to $\pm 1$, we have that
\ba
E^{(0)}_{\left|\alpha\right>} &=& -(\text{\# of terms satisfied in }\alpha) + (\text{\# unsat. in }\alpha) \nn
&=&-2(\text{\# of terms satisfied in }\alpha)+m
\ea

Then, because $\left|\alpha\right>$ is a uniform superposition over the states corresponding with $\alpha$, the $1^{\text{st}}$-order perturbative energy correction of $\left|\alpha\right>$ is
\ba
E^{(1)}_{\left|\alpha\right>} &=& \left<\alpha\right|H_B\left|\alpha\right> \nn
&=& -\sum_{i\text{ degen. in } \alpha} \Delta_i \nn
&=&-a(\text{\# of terms satisfied in }\alpha) \nn
&=&{a \over 2}\left(E^{(0)}_{\left|\alpha\right>} - m\right) \nn
&=&{a \over 2}E^{(0)}_{\left|\alpha\right>} - \text{constant}.
\ea

This means that a final state $\alpha$ of lower energy than a final state $\beta$, will have a lowest eigenstate $\left|\alpha\right>$ with larger negative slope than $\left|\beta\right>$. Thus, $1^{\text{st}}$-order perturbation predicts that $\left|\alpha\right>$ and $\left|\beta\right>$ will diverge and not cross.

In the case where $\alpha$ is degenerate in the original Ising model, (but with no 1 bit flip degeneracy), it can be seen that $\left|\alpha\right>$ is a combination of uniform superpositions over each state's new degeneracy.  For example, if $\alpha$ originally contained states $k$ and $l$, $\left|\alpha\right>$ contains some combination of a uniform superposition over the states corresponding with $k$ and a uniform superposition over the states corresponding with $l$.  Since there are assumed to be no $1^{\text{st}}$-order degeneracies in the original $\alpha$, this means that $E^{(1)}_{\left|\alpha\right>}$ is the same as in the non-degenerate case.

However, in order to prove that this can eliminate perturbative crossings, one must examine the impact of this construction on higher orders of perturbation.  Consider a single state $\alpha_*$ in the eigenstate $\left|\alpha\right>$.  If $\alpha_*$ is a local minimum (a state with which a crossing could occur with the global minimum), all states 1 bit flip from $\alpha_*$ have higher energy, and therefore have at least one $h$ or $J$ unsatisfied that is satisfied in $\alpha_*$.  As decribed above, for any $h$ or $J$ satisfied in $\alpha_*$, each of the corresponding $a$ added qubits contributes half of the states in the new superposition $\left|\alpha_*\right>$.  This means that flipping any 1 bit, $i$, from $\alpha_*$, which had incurred energy cost $B_{\alpha_*,i}$, now incurs at least cost $(1+2kb)B_{\alpha_*,i}$ for the portion of $\left|\alpha_*\right>$ that has $k$ added qubits in their $+1$ states.  Since the choice of $\alpha_*$ was arbitrary, this applies to all of $\left|\alpha\right>$.  For simplicity of analysis, we choose $a=b=n^2\rightarrow \infty$, though this is extremely excessive in practice.  We then find that the $2^{\text{nd}}$ order energy correction is bounded by

\ba
&&\left|E^{(2)}_{\left|\alpha\right>}\right| \nn
&\le& \left|E^{(2)}_{\alpha \text{ orig.}}\right|{1 \over 2^a}\sum_{k=0}^{a}\binom{a}{k}(1+2kb)^{-1} + \sum_{\substack{\text{added } i\\ \text{ nondegen. in }\alpha}} {\Delta_i^2 \over 2b} \nn
&\rightarrow& \left|E^{(2)}_{\alpha \text{ orig.}}\right|{1 \over ab} + {a \over 2b}(\text{\# of terms unsatisfied in }\alpha) \nn
&=& \left|E^{(2)}_{\alpha \text{ orig.}}\right|{1 \over n^4} + {1 \over 4}\left(E^{(0)}_{\left|\alpha\right>}+m\right) \nn
&=& O(n^{-2}) + {1 \over 4}\left(E^{(0)}_{\left|\alpha\right>}+m\right) \nn
&\rightarrow& {1 \over 4}E^{(0)}_{\left|\alpha\right>} + \text{constant}.
\ea

One may wonder why $a=b$ was chosen instead of $b\gg a$, since the latter would make the $2^{\text{nd}}$ order term approach zero, instead of favouring higher-energy final states as it currently does.  That, however, could also make higher-order terms approach zero in a similar manner, dramatically increasing the radius of convergence of the series.  By setting $a=b$ and expanding a component of the perturbation, one finds that for $q\ge 2$, the $q$th-order term will have a component of magnitude $(E^{(0)}_{\left|\alpha\right>}/4)^{q-1}$ (of alternating sign) and no components proportional to $a$.  For example, it is easily checked that the $3^{\text{rd}}$ order term is dominated by $(1/16)(E^{(0)}_{\left|\alpha\right>})^2$.  This means that the radius of convergence scales proportional to $1/E^{(0)}_{\left|\alpha\right>}$, getting smaller as the problem size increases.  Since the magnitude of the $1^{\text{st}}$ order term is proportional to $a$, and the value of $a$ was chosen to be $\Theta(n^2)$, within the radius of convergence, $1^{\text{st}}$ order should dominate $\Omega(n^2 E^{(0)}_{\left|\alpha\right>})$ orders of perturbation.

Choosing such a large $a$ also guarantees elimination of crossings if $\textcolor{darkblue}{M}$ had $1^{\text{st}}$ order degeneracy of its local minima, since for any $\textcolor{darkblue}{M}$ with at least one non-zero term, $E^{(1)}_{\alpha \text{ orig.}} > -n$.  After adding the extra qubits, the difference between the $1^{\text{st}}$ order terms will still be $\Omega(n^2)$, in favour of the global minimum.

One may argue that adding qubits with such a large factor $b$ could simply add a new crossing at a point where $\lambda \sim b$ and the original diagonal part of the Hamiltonian can be neglected, effectively giving a diagonal part of just
\ba
\sum_{i\in \textcolor{darkblue}{M}}\sum_{k=1}^{a} \textcolor{darkgreen}{b (h_i \sigma_z^{(i)}+1)(\sigma_z^{(i_k)}+1)/2} + \nn \sum_{\{i,j\}\in \textcolor{darkblue}{M}}\sum_{k=1}^{a} \textcolor{darkgreen}{b (J_{ij}\sigma_z^{(i)}\sigma_z^{(j)}{+}1) (\sigma_z^{(ij_k)}{+}1)/2}.
\ea
However, the ground state of this Hamiltonian is exponentially degenerate and one of these states is always trivially found.  This is because there are no local minima: any nonzero term can be made zero by flipping the corresponding added qubit ($i_k$ or $ij_k$) from $+1$ to $-1$.

\section{Extending to arbitrary coefficients}
The above construction for Ising model terms with coefficients in $\{-1,0,+1\}$ is most easily extended to integer coefficients.  For a term with integer coefficient $+k$, simply rewrite it as $k$ identical terms with coefficient $+1$ and apply the construction as before on all $k$ terms.  The same splitting works with coefficients of $-k$ becoming $k$ terms with coefficient $-1$.

To extend this beyond integers to polynomial precision real numbers, one could simply rescale $\textcolor{darkblue}{M}$ to be polynomially approximated by integers, but a more reasonable construction is available from that
\ba
E^{(1)}_{\left|\alpha\right>} &=& -\sum_{i\text{ degen. in } \alpha} \Delta_i.
\ea

One can simply round a coefficient down to the nearest integer below, apply the construction as before, then add $a$ extra qubits as above whose $\Delta_i$ values are selected to be the remainder that was rounded off, instead of $1$.  This has negligible impact on higher order terms in that the radius of convergence is still proportional to $1/E^{(0)}_{\left|\alpha\right>}$, (unless the coefficients were chosen to be scaled such that the bulk of weight lies in the rounded off components, which is easily fixed by scaling up the coefficients into a more reasonable range).

\section{Conclusion}
\label{Conclusion}
Above, we have presented a simple method of posing the NP-hard problem of finding the ground state of an arbitrarily-connected Ising model with local fields as an adiabatic quantum optimization with no perturbative crossings between local and global minima.  Thus, all NP-complete problems can be solved using adiabatic quantum optimization without encountering these crossings.

It is critical to note that this does {\em not} prove that adiabatic quantum optimization can solve NP-complete problems in polynomial time.  However, it does mean that proving otherwise requires identifying some effect other than perturbative crossings that unavoidably results in exponentially long adiabatic runtimes.

\section*{Acknowledgements}

We thank M.H.S.~Amin for the coaxing to come up with the idea behind this work and for many useful discussions.  We also thank I.~Affleck, E.~Farhi, F.~Hamze, K.~Karimi, H.~Katzgraber, R.~Raussendorf, and A.P.~Young for useful discussions, as well as M. Johnson, T. Lanting, T. Mahon, C. Rich, M. Thom, and B. Wilson for proofreading assistance.

\section{Appendix: Reduction to 2-Local}
\label{Reduction}
As mentioned above, the prescription above introduces a 3-local term when adding an extra qubit corresponding to $J$ terms in the original Ising model.  However, a 2-local function with the necessary properties can be constructed by adding another extra qubit.  The properties to maintain are:
\begin{enumerate}
\item all values of the function are $\ge 0$
\item all configurations of qubits $i$ and $j$ that satisfy the associated $J_{ij}\ne 0$ term must have 2 configurations of the additional qubits, 1 bit flip apart, where the function is 0
\item all configurations of qubits $i$ and $j$ that do not satisfy the associated $J_{ij}\ne 0$ term must have 1 configuration of the additional qubits where the function is 0
\item all other configurations must give values $\ge 2b$.
\end{enumerate}

The following 4-qubit, 2-local function satisfies these properties for a positive $J$.  $(\sigma_z^{(i)}+1)/2$ is abbreviated as $x_i$, etc.
\ba
&&f_{(+)}(\sigma_z^{(i)},\sigma_z^{(j)},\sigma_z^{(ij_k)},\sigma_z^{(ij_k*)}) \nn
&=& {b\over 2} \Big( 4\left(x_i x_j + x_j x_{ij_k} + x_{ij_k} x_i\right) \nn
&& +6\left(x_i + x_j + x_{ij_k}\right)(1-2x_{ij_k*}) + 8x_{ij_k*} - 1 \nn
&& +\sigma_z^{(i)}\sigma_z^{(j)} + \sigma_z^{(ij_k)} + 1\Big)
\ea

The first 2 lines inside the outer parentheses encode a $+\sigma_z^{(i)}\sigma_z^{(j)}\sigma_z^{(ij_k)}$ term.  Note that alternate encodings of this 3-local term where exactly 2 of $x_i$, $x_j$, and $x_{ij_k}$ are flipped (e.g. replaced with $1-x_i$ etc.) do not maintain the needed properties of the overall function.  The cost table of $f_{(+)}$ is as follows:

\begin{center}
\begin{tabular}{| c | c | c || c  c |}
\hline
\multicolumn{3}{|c||}{ } & \multicolumn{2}{c|}{$f_{(+)}/2b$} \\
\hline
$\sigma_z^{(i)}$ & $\sigma_z^{(j)}$ & $\sigma_z^{(ij_k)}$ & \multicolumn{1}{c|}{$\sigma_z^{(ij_k*)}=-1$} & $\sigma_z^{(ij_k*)}=+1$ \\
\hline
$-1$ & $-1$ & $-1$ & {\bf 0} & 2 \\
$-1$ & $-1$ & +1 & 2 & 1 \\
\hline
$-1$ & +1 & $-1$ & 1 & {\bf 0} \\
$-1$ & +1 & +1 & 4 & {\bf 0} \\
\hline
+1 & $-1$ & $-1$ & 1 & {\bf 0} \\
+1 & $-1$ & +1 & 4 & {\bf 0} \\
\hline
+1 & +1 & $-1$ & 4 & {\bf 0} \\
+1 & +1 & +1 & 8 & 1 \\
\hline
\end{tabular}
\end{center}

It's clearly visible that for some value of new qubit $ij_k*$, qubit $ij_k$ is at degeneracy when the original $J$ term is satisfied, but when the $J$ term is unsatisfied, neither qubit is at degeneracy.

The following 4-qubit, 2-local function satisfies these properties for a negative $J$:
\ba
&&f_{(-)}(\sigma_z^{(i)},\sigma_z^{(j)},\sigma_z^{(ij_k)},\sigma_z^{(ij_k*)}) \nn
&=& {b\over 2} \Big( 4\left((1-x_i) x_j + x_j x_{ij_k} + x_{ij_k} (1-x_i)\right) \nn
&& +6\left((1-x_i) + x_j + x_{ij_k}\right)(1-2x_{ij_k*}) + 8x_{ij_k*} - 1 \nn
&& -\sigma_z^{(i)}\sigma_z^{(j)} + \sigma_z^{(ij_k)} + 1\Big)
\ea

The first 2 lines inside the outer parentheses encode a $-\sigma_z^{(i)}\sigma_z^{(j)}\sigma_z^{(ij_k)}$ term.  Instead of flipping $x_i$ (i.e. replacing with $1-x_i$), it also works to instead flip either $x_j$ or $x_{ij_k}$, so long as it is only 1 of the 3.  Note that an alternate encoding of this 3-local term where all 3 are flipped does not maintain the needed properties of the overall function.  The cost table of $f_{(-)}$ as defined above is as follows:

\begin{center}
\begin{tabular}{| c | c | c || c  c |}
\hline
\multicolumn{3}{|c||}{ } & \multicolumn{2}{c|}{$f_{(-)}/2b$} \\
\hline
$\sigma_z^{(i)}$ & $\sigma_z^{(j)}$ & $\sigma_z^{(ij_k)}$ & \multicolumn{1}{c|}{$\sigma_z^{(ij_k*)}=-1$} & $\sigma_z^{(ij_k*)}=+1$ \\
\hline
$-1$ & $-1$ & $-1$ & 1 & {\bf 0} \\
$-1$ & $-1$ & +1 & 4 & {\bf 0} \\
\hline
$-1$ & +1 & $-1$ & 4 & {\bf 0} \\
$-1$ & +1 & +1 & 8 & 1 \\
\hline
+1 & $-1$ & $-1$ & {\bf 0} & 2 \\
+1 & $-1$ & +1 & 2 & 1 \\
\hline
+1 & +1 & $-1$ & 1 & {\bf 0} \\
+1 & +1 & +1 & 4 & {\bf 0} \\
\hline
\end{tabular}
\end{center}

It's again visible that qubit $ij_k$ is at degeneracy when and only when the original $J$ term is satisfied.

\end{document}